\documentclass[reprint,longbibliography,preprintnumbers,nofootinbib,amsmath,amssymb,aps,onecolumn]{revtex4-2}
\usepackage{dblfloatfix}
\usepackage{graphicx}% Include figure files
\usepackage{dcolumn}% Align table columns on decimal point
\usepackage{bm}% bold math
\usepackage{color}
\usepackage{amssymb}
\usepackage{pifont}
\definecolor{green2}{rgb}{0.0, 0.5, 0.0}
\usepackage[dvipsnames]{xcolor}
\usepackage[colorlinks = true,
            linkcolor = blue,
            urlcolor  = blue,
            citecolor = green,
            anchorcolor = blue]{hyperref}
\usepackage{verbatim}
\makeatletter
\usepackage{multirow, graphicx,amssymb,url,mathrsfs,amsmath}
\usepackage{amsxtra,amstext,latexsym,dsfont,amsfonts}
\usepackage{color,eucal}
\usepackage[dvipsnames]{xcolor}
\usepackage{float}
\usepackage{colortbl}
\usepackage{multirow}
\usepackage{tikz}
\usetikzlibrary{tikzmark}
\usetikzlibrary{arrows}

\newcommand{\Rmnum}[1]{\expandafter\@slowromancap\romannumeral #1@}
\usepackage{tabularx, array}
\newcolumntype{L}[1]{>{\raggedright\arraybackslash}p{#1}}
\newcolumntype{C}[1]{>{\centering\arraybackslash}p{#1}}
\newcolumntype{R}[1]{>{\raggedleft\arraybackslash}p{#1}}
\makeatother

\makeatletter
\newcommand{\myBig}{\bBigg@{1.75}}
\makeatother

\begin{document}
\title{\Large \textbf{\textbf{Holographic s+p superconductors with nonlinear electrodynamics} } }
\author{Ru-Qing Chen}
\affiliation{Center for gravitation and astrophysics, Kunming University of Science and Technology, Kunming 650500, China}
\author{Hui Zeng}
\email{zenghui@kust.edu.cn (corresponding author)}
\affiliation{Center for gravitation and astrophysics, Kunming University of Science and Technology, Kunming 650500, China}
\author{Zhang-Yu Nie}
\email{niezy@kust.edu.cn (corresponding author)}
\affiliation{Center for gravitation and astrophysics, Kunming University of Science and Technology, Kunming 650500, China}
\date{\today}
\begin{abstract}
We investigate a holographic s+p superconductor model coupled to nonlinear electrodynamics in the probe limit. The equations of motion are solved numerically, and the condensates as well as the grand potential curves for phase transitions at different values of the nonlinear parameter $b$ are illustrated. It is found that as $b$ increases, both the pure s-wave and p-wave condensates are suppressed. From the $b-T$ phase diagram, we observe that the region of the pure s-wave phase gradually shrinks with increasing $b$, which is attributed to the stronger suppression on the s-wave condensate compared to the one on the p-wave. Moreover, a smaller charge ratio $q_p/q_s$ is needed for the s+p coexistent phase to appear as $b$ grows. A particularly interesting feature is that, due to the nonlinear self-interaction of the electromagnetic field, charge accumulates spontaneously outside the event horizon from the bulk perspective even in the absence of scalar and vector condensates, thereby invalidating the conventional formula for the superconducting charge density. We improve the definition of the superconducting charge density as the background-subtracted value of the accumulated charge outside the horizon with respect to that in the normal phase. Furthermore, the optical conductivity in the normal phase is also modified by this accumulated charge for finite $b$, and its imaginary part develops a minimum at a finite frequency. This minimum persists in the superconducting phase near the critical point, confusing the extraction of the gap frequency.
\end{abstract}
\maketitle
%\tableofcontents
%\tableofcontents
\section{Introduction}\label{sec1}
The gauge/gravity duality, also known as the AdS/CFT correspondence \cite{Maldacena:1997re}, is a bridge connecting strongly coupled field theories with weakly coupled gravitational theories and therefore attracted extensive attention in high-energy physics and condensed matter physics, especially after the holographic modeling of superconductors in the HHH model \cite{Gubser:2008px,Hartnoll:2008vx} proposed by Hartnoll, Herzog, and Horowitz. In addition to the s-wave order introduced in the HHH model, p-wave and d-wave orders are also included in the holographic studies \cite{Gubser:2008wv,Chen:2010mk}, which provide new perspectives to understand superconductors with different symmetries. Subsequently, the holographic superconductor models with with multi-condensate, such as the s+s models~\cite{Basu:2010fa,Cai:2013wma,Li:2017wbi,Zhang:2025hkb}, the s+p models~\cite{Nie:2013sda,Nie:2014qma,Amado:2013lia,Nie:2015zia,Arias:2016nww,Xia:2021pap,Zhang:2023uuq,Zhang:2025tsa,Chen:2025kmo} as well as the s+d models~\cite{Nishida:2014lta,Li:2014wca,Liao:2025knf}, are also realized to study the competition and co-existence between the various order parameters, which is natural to be considered in a complex real system~\cite{Huang:2011ac,Krikun:2013hwi,Musso:2013ija,Donos:2012yu,Wen:2013ufa} and makes the phase transitions more abundant.

In the recent studies on holographic superconductor models, nonlinear potential terms of the scalar and vector fields are applied to realize more accurate control on the phase transitions~\cite{Herzog:2010vz,Zhao:2022jvs,
Zhao:2023ffs,Zhao:2024jhs,Cao:2024irr,Zhao:2025ecg,Wang:2025ajn,Zhao:2025odj}. Meanwhile, Nonlinear models for the gauge fields such as Born-Infeld electrodynamics (BINE)~\cite{Born:1934gh} is also introduced in the study of holographic superconductors~\cite{Jing:2010zp}. Moreover, other types of nonlinear electrodynamics have also been proposed, such as logarithmic nonlinear electrodynamics (LNE)~\cite{Soleng:1995kn}, exponential nonlinear electrodynamics (ENE)~\cite{Hendi:2012zz}, and nonlinear electrodynamics with a simple nonlinear term ($bF^4$)~\cite{Mohammadi:2019swg}. It is also found that nonlinear electrodynamics help avoid singularities inside black holes~\cite{Ayon-Beato:1998hmi}. Although holographic superconductors coupled to nonlinear electrodynamics have been widely studied in the context of the single s-wave or p-wave order\cite{Gangopadhyay:2012am,Zhao:2012cn,Liu:2015lit,Lai:2015rva,Sheykhi:2015mxb,Sheykhi:2016aoi,Sheykhi:2016kqh,Sheykhi:2016aoi,Sheykhi:2016meb,Sheykhi:2018mzs,Ghotbabadi:2018ahu,Mohammadi:2019ndh,Mohammadi:2020jbe,Yao:2021cns,Mohammadi:2022kkv,Calderon:2024fuv,Wei:2025yrl}, the competition and coexistence between the two different orders have not yet been investigated. It is therefore interesting to further investigate how the nonlinear electrodynamics changes the phase transitions, the superconducting charge density, and the optical conductivity in a model with both the s-wave and p-wave orders.

In this study, we construct a holographic s+p superconductivity model coupled to nonlinear electrodynamics. We consider a simplest version of nonlinear electrodynamics with the form $\mathcal{L}(F)=\mathcal{F}+b\mathcal{F}^2$, where $\mathcal{F}= -\frac{1}{4}F_{\mu\nu}F^{\mu\nu}$ is the standard Maxwell Lagrangian density, and b is the nonlinear parameter. Such a form is the simplest higher-order correction to Maxwell electrodynamics that still preserves gauge invariance, and give the universal lowest level corrections for general nonliear electrodynamics. We will investigate the competition and coexistence between the s-wave and p-wave orders in the probe limit for different values of the nonlinear parameter b, as well as study the charge density and optical conductivity for the various superconductor phases.

The plan of the rest of this paper is as follows. In Section~\ref{sec2}, we establish a holographic s+p superconductor model with nonlinear electrodynamics in the probe limit. In Section~\ref{sec3}, we investigate the impact of the nonlinear parameter b on the phase structure. 
In Section~\ref{sec4}, we study the superconducting charge density. In Section ~\ref{sec5}, we study the optical conductivity to show the dependence of the energy gap on the nonlinear parameter b. In Section~\ref{sec6}, we provide conclusions and discussions.
\section{The holographic s+p model coupled to nonlinear electrodynamics} \label{sec2}
We work in the probe limit and consider the asymptotically AdS black hole with a planar horizon as the background, which is described by the following line element
\begin{align}
&\ ds^2 =-f(r)dt^2+\frac{dr^2}{f(r)}+\frac{r^2}{L^2}(dx^2 + dy^2),
\end{align}
where
\begin{align}
&\ f(r)=\frac{r^2}{L^2}\left(1-\frac{r_h^3}{r^3}\right)~.
\end{align}
In this line element, $L$ is called the AdS radius and $r_h$ denotes the location of the event horizon with $f(r_h)=0$. The Hawking temperature is given by
\begin{align}
T= \frac{3 r_h}{4\pi L^2}~.
\end{align}

The action of the matter fields of the holographic s+p model coupled to nonlinear electrodynamics is
\begin{align}
S_M=&\int d^{4}x\sqrt{-g}\Big(\mathcal{L}(F)
-D_{\mu}\Psi^{\ast}D^{\mu}\Psi-m_s^{2}\Psi^{\ast}\Psi-\frac{1}{2} \rho_{\mu\nu}^{\dagger} \rho^{\mu\nu}-m_p^2\rho^\dagger_\mu \rho^\mu\Big)~,\label{Lagm}
\end{align}
where
\begin{align} 
\mathcal{L}(F)=\mathcal{F}+b\mathcal{F}^2~,
\end{align}
with
\begin{align}
\mathcal{F}= -\frac{1}{4}F_{\mu\nu}F^{\mu\nu}~.
\end{align}
When the nonlinear parameter $b$ equals $0$, the system returns to the s+p model coupled to the linear Maxwell electrodynamics. $F_{\mu\nu}=\nabla_{\mu}A_{\nu}-\nabla_{\nu}A_{\mu}$ is the Maxwell field strength, $\Psi$ is the complex scalar field, and $\rho_\mu$ is the complex vector fields with the field strength $\rho_{\mu\nu}=\bar{D}_\mu\rho_\nu-\bar{D}_\nu \rho_\mu$. The covariant derivatives are given by $D_{\mu}\Psi=\nabla_{\mu}\Psi-i q _sA_\mu\Psi$ and $\bar{D}_\mu \rho_{\nu}=\nabla_\mu\rho_{\nu}-iq_p A_\mu\rho_{\nu}$, respectively. $q_s$, $q_p$ and $m_s$, $m_p$ are the charges and masses of $\Psi$ and  $\rho_\mu$, respectively. 
 
We set the following ansatz for the matter fields
\begin{align}\label{ansatz}
\quad A_t =\Phi(r)~, \quad\Psi=\Psi_s(r)~,\quad\rho_x=\Psi_p(r) ~,
\end{align}
and all other field components are set to zero. Consequently, the equations of motion for these matter fields are as follows 
\begin{align}  
    \left[\frac{2 q_p^2 L^2 \Psi_p^2(r)+2 q_s^2 r^2 \Psi_s^2(r)}{r^2 f(r) (1 + 3b \Phi'^2(r))} \right]\Phi(r)-\frac{2}{r} \left[ \frac{1 + b \Phi'^2(r)}{1 + 3b \Phi'^2(r)} \right] \Phi'(r)-\Phi''(r) = 0,~\label{equation11}\\   
    \frac{m_s^2 \Psi_s}{f} - \frac{q_s^2 \Phi^2 \Psi_s}{f^2} - \frac{2 \Psi_s'}{r} - \frac{f' \Psi_s'}{f} - \Psi_s'' = 0,~\label{equation12}\\
    \frac{m_p^2 \Psi_p}{f} - \frac{q_p^2 \Phi^2 \Psi_p}{f^2} - \frac{f' \Psi_p'}{f} - \Psi_p'' = 0.~\label{equation13}
\end{align}

To solve these coupled equations, we need to specify the boundary conditions. The expansions near the horizon $r=r_{h}$ are
\begin{align}
\Phi(r) &= \Phi_1(r - r_h) + O(r - r_h)^2~,  \\
\Psi_s(r) &= \Psi_{s0} + \Psi_{s1}(r - r_h) + O(r - r_h)^2~, \\
\Psi_p(r) &= \Psi_{p0} + \Psi_{p1}(r - r_h) + O(r - r_h)^2~.
\end{align}
The value of $\Phi(r=r_h)$ is set to zero to ensure the finiteness of $g^{\mu\nu}A_{\mu}A_{\nu}$.

The asymptotic expansions near the boundary $r\to\infty$ are given by
\begin{align}
\Phi(r) &= \mu - \frac{\rho}{r} + \ldots,  \\
\Psi_s(r) &= \frac{\Psi_{s-}}{r^{\Delta_{s^-}}} + \frac{\Psi_{s+}}{r^{\Delta_{s^+}}} + \ldots,\\
\Psi_p (r)&= \frac{\Psi_{p-}}{r^{\Delta_{p^-}}} + \frac{\Psi_{p+}}{r^{\Delta_{p^+}}} + \ldots, 
\end{align}
where
\begin{align}
\Delta_{s\pm} &= \frac{3\pm \sqrt{9 + 4m_s^2}}{2},  \\
\Delta_{p\pm} &= \frac{1 \pm \sqrt{1 + 4m_p^2}}{2}.
\end{align}
Due to the AdS/CFT dictionary, $\mu$ and $\rho$ are the chemical potential and charge density of the boundary system, respectively. We adopt the standard quantization, in which $\Psi_{s-}$ and $\Psi_{p-}$ are considered the source terms of the boundary operators, while $\Psi_{s+}$ and $\Psi_{p+}$ are regarded as the vacuum expectation values. We set the source free boundary conditions $\Psi_{s-}=\Psi_{p-}=0$ to obtain the solutions for spontaneous U(1) symmetry breaking.

In this paper, we work in the grand canonical ensemble. To compare the stability of the various solutions, we calculate the grand potential to determine the stability relationships between them. From the AdS/CFT duality, the grand potential equals to the on shell Euclidean bulk action. In the probe limit, the contributions from the gravity part are the same for different solutions. Therefore the difference in the grand potential originates solely from the matter part of the action. After substituting the equations of motion into the matter part of the Euclidean action, the contribution to the grand potential from the matter part $\Omega_m$ is
%The $\Omega_m$ represents the contribution from the matter action to the grand potential. 
\begin{align}
\Omega_m=\frac{V_2 }{T}(-\frac{\mu\rho }{2 L^2}+\int_{r_h}^{\infty}( \frac{1}{4 L^2} b r^2 \Phi'(r)^4 -\frac{\Phi(r)^2 \left(  L^2q_p^2 \Psi_p(r)^2 + q_s^2 r^2 \Psi_s(r)^2\right)}{ L^2f(r) \left(1 + 3 b \Phi'(r)^2\right)} -\frac{2 b r \Phi(r) \Phi'(r)^3}{  L^2(1 + 3 b \Phi'(r)^2)})dr)~.
\end{align}

The equations of motion (\ref{equation11},\ref{equation12},\ref{equation13}) exhibit the following three sets of scaling symmetries, which are useful to simplify the numerical procedures.
\begin{align}
(\text{I})~ &\Phi \rightarrow \lambda \phi, \Psi_p \rightarrow \lambda\Psi_p, f \rightarrow \lambda^{2} f, r \rightarrow \lambda r,  r_h \rightarrow \lambda r_h~; 
\label{scaling1}\\
(\text{II})~ &\Phi \rightarrow \lambda^{-2} \phi,\Psi_p \rightarrow \lambda^{-2}\Psi_p,  \Psi_s \rightarrow \lambda^{-1}\Psi_s,f \rightarrow \lambda^{-2}f,  ~\nonumber\\
&m_s^2\rightarrow \lambda^{-2}m_s^2,m_p^2\rightarrow \lambda^{-2}m_p^2,  L \rightarrow \lambda L,b \rightarrow \lambda^{4}b~; 
\label{scaling2}\\
(\text{III})~ &\Phi \rightarrow \lambda^{-1} \phi, \Psi_p \rightarrow \lambda^{-1}\Psi_p,\Psi_s \rightarrow \lambda^{-1}\Psi_s,  \nonumber\\
&\ q_s \rightarrow \lambda q_s, \ q_p \rightarrow \lambda q_p,b \rightarrow \lambda^{2}b~.
\label{scaling3}
\end{align}
Applying the scaling symmetries (\ref{scaling1}) and (\ref{scaling2}), $r_h$ and $L$ can be rescaled to any value. $q_s$ and $q_p$ can also be rescaled by the scaling symmetry (\ref{scaling3}), while the ratio $q_p/q_s$ is stable. Therefore, in the rest of this paper, we set $L=r_h=q_s=1$ without loss of generality. To focus on the effects of the nonlinear parameter b , we fix the mass parameters as $m_s^2 = 0$ and $m_p^2 =3/4$ throughout the remainder of this paper.
\section{The phase transitions involving the s-wave and p-wave orders} \label{sec3}
\subsection{The single condensate s-wave and p-wave solutions}\label{sec3.1}
\begin{figure}
\includegraphics[width=0.39\columnwidth]{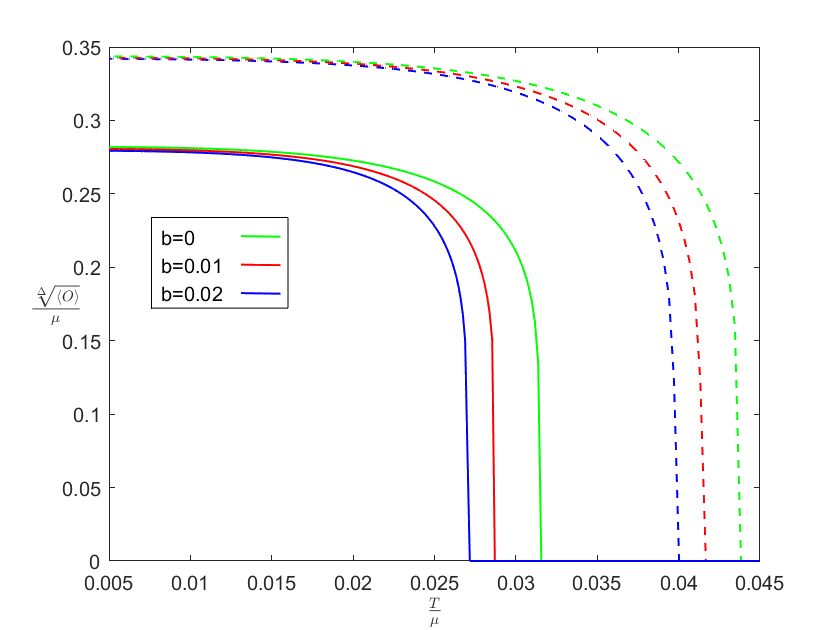}
\includegraphics[width=0.39\columnwidth]{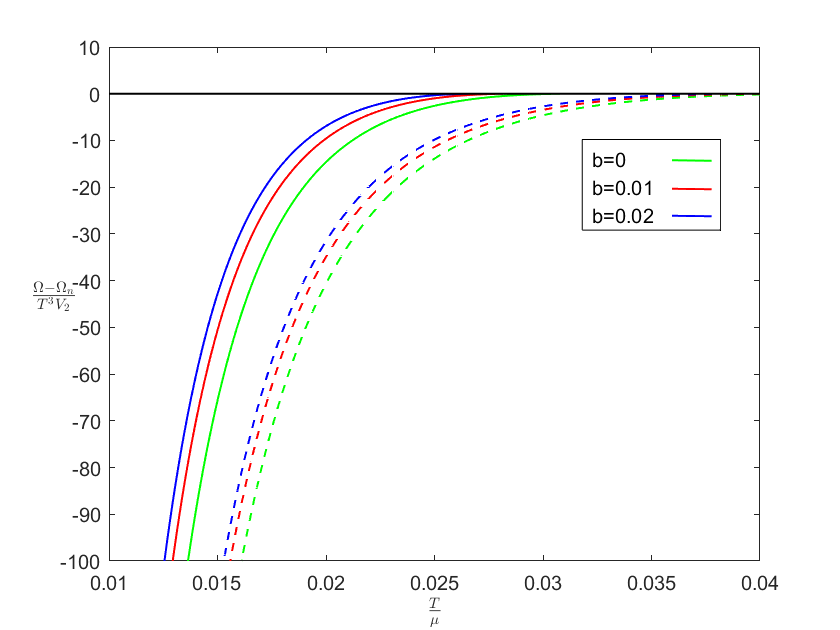} 
\caption{The condensate(Left Plot) and grand potential curves (Right Plot) of the s-wave (solid lines) and p-wave (dashed lines) solutions are presented for $b$=0 (green), 0.01 (red), and 0.02 (blue). %Left: condensate curves of the s-wave (solid) and p-wave (dashed) solutions. Right:  grand-potential curves of the s-wave (solid) and p-wave (dashed) solutions.
}\label{1}
\end{figure}

In this system, there are three distinct solutions with non-zero condensates: the s-wave solution, the p-wave solution, and the s+p coexistent solution. By turning off either the p-wave or s-wave order, we obtain the s-wave solution or the p-wave solution. In Figure~\ref{1}, we plot the condensate and grand potential density curves of the single condensate s-wave and p-wave solutions with $q_p=1$ and three different values of the nonlinear parameter $b$. We can see that as b increases, the single condensate s-wave and p-wave solutions exhibit the same qualitative law: their critical values of temperature $T_c/\mu$ both increase, and their grand potential curves both rise up, indicating a suppression on the single condensate solutions. Meanwhile, the condensate value of the s-wave and p-wave solutions in the low temperature limit both decrease as the nonlinear parameter $b$ increases~\cite{Mohammadi:2019swg,Mohammadi:2022kkv}.
\subsection{The multi-condensate s+p solutions}\label{sec3.2}
\begin{figure}
\includegraphics[width=0.39\columnwidth]{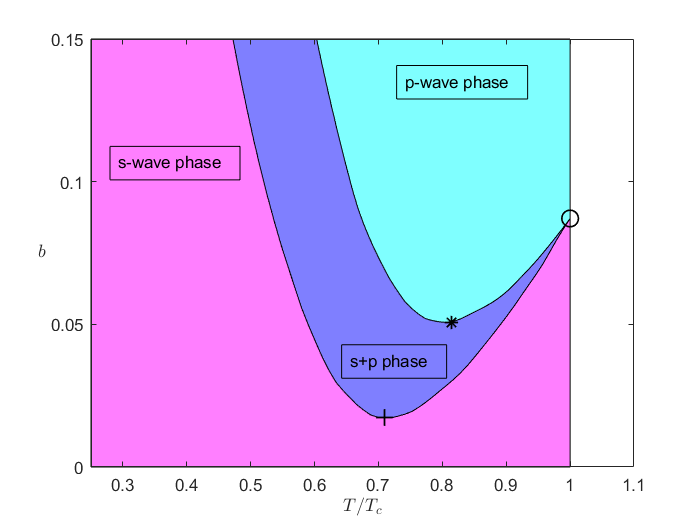}
\caption{The $b-T$ phase diagram with $q_p=0.7183$. The white, magenta, cyan and blue regions are dominated by the normal phase, the s-wave phase, the p-wave phase and the s+p phase, respectively. The symbols \textit{$+$}, \textit{$*$}, \textit{$o$} are used to mark the three special points who's value of $b$ are denoted as $b^*$, $b^+$ and $b^o$, respectively.}\label{2}
\end{figure}

In order to present the competition and coexistence between the s-wave and p-wave orders, we need to also get the s+p coexistent solutions and compare the grand potential for the various solutions. We first choose $q_p/q_s=q_p=0.7183$ and focus on the influence of the nonlinear parameter $b$. In Figure~\ref{2}, we plot the $b-T$ phase diagram to present the overall results, where four distinct regions are observed, which indicate the normal phase (white), the s-wave phase (magenta), the p-wave phase (cyan), and the multi-condensate s+p phase (blue), respectively. The boundary lines between these regions represent second-order critical points. From this phase diagram, we can see that when $b=0$, only the s-wave order condensates in the low temperature region. As the value of $b$ increases and reaches the special value $(b)^+$, the s+p phase emerges from the middle region of the s-wave phase. The region of the s+p phase grows while the region of the s-wave phase shrinks along the increasing of $b$. When b reaches the second special value $(b)^*$, a single condensate p-wave phase begin to dominate between two sections of s+p phases. There is a third special value $b^o$ that corresponds to the quadrupole point. When $b$ is larger than $b^o$, the single condensate p-wave phase dominates the higher temperature region while the single condensate s-wave phase dominates the low temperature region, between them exhibits a single section of the typical ``X-type'' s+p coexistent phase. We have also confirmed that at sufficiently large value of $b$, only a stable single p-wave solution dominates. Since the stability of both the single condensate s-wave and p-wave solutions are reduced with the increasing of $b$, this phase diagram further indicates that the reduction of the stability for the single condensate p-wave solution is slower than that of the single condensate s-wave solution.

\begin{figure}
\includegraphics[width=0.3\columnwidth]{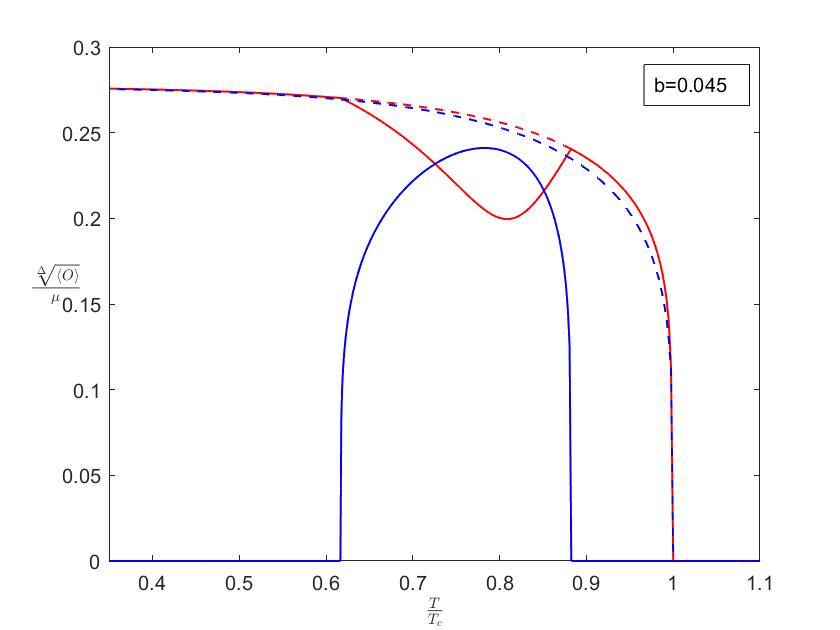}
\includegraphics[width=0.3\columnwidth]{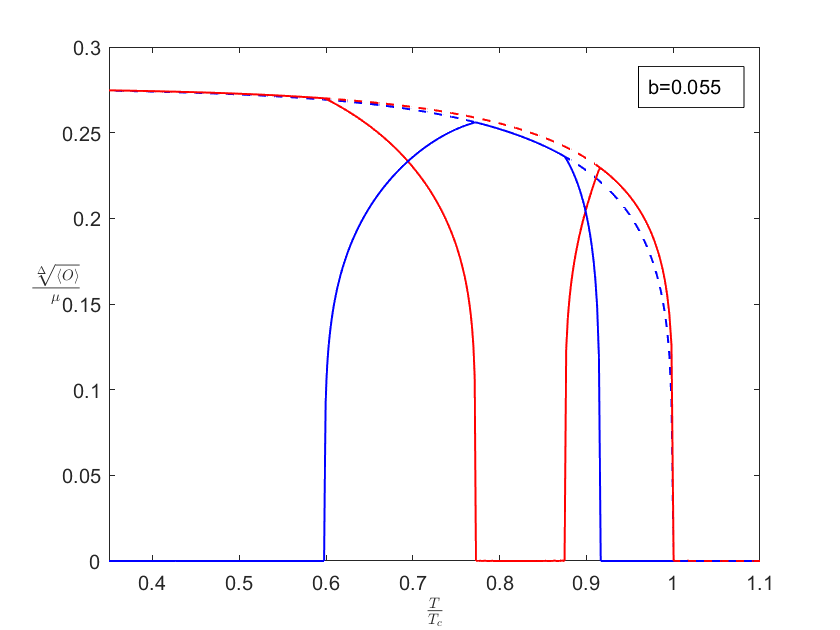}
\includegraphics[width=0.3\columnwidth]{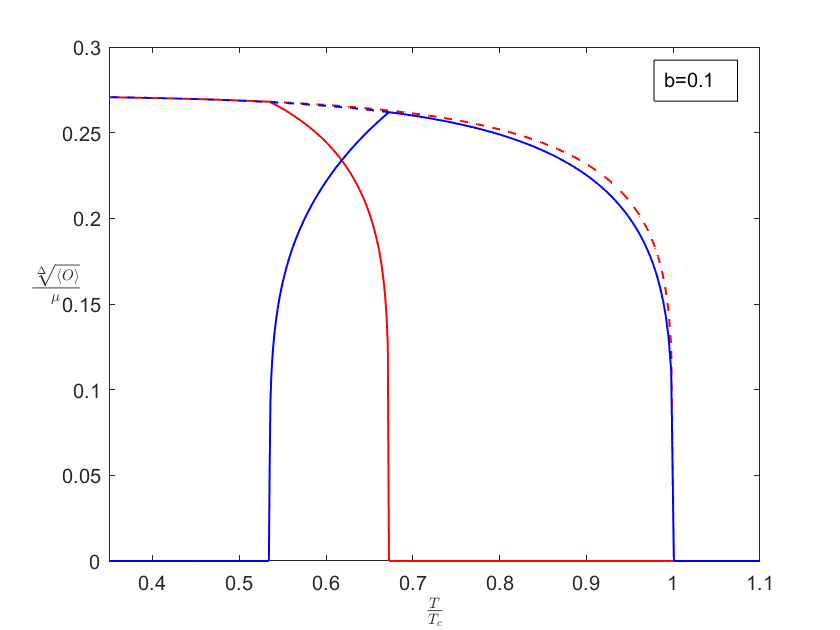}  
	\caption{The condensate curves for the three typical cases with $b=0.045$, $0.055$ and $0.1$, respectively. The red and blue lines indicate the condensate values of the s-wave and p-wave orders, respectively. Solid lines denote the condensate values for the most stable solutions, while dashed lines denote the condensate values for the unstable sections of the single condensate solutions.
}\label{3}
\end{figure}

In Figure~\ref{3}, we further present the detailed condensate curves involving the competition and coexistence between the s-wave and p-wave orders for three typical values of $b$. In these plots, the red and blue lines represent the s-wave and p-wave orders, respectively. The solid lines denote the condensate values for the most stable solutions, while dashed lines denote the condensate values for the unstable sections. From the phase diagram we can see that when $b<b^+$, only the s-wave order condensates in the low temperature region. For the first case involving both the s-wave and p-wave orders with $b^+<b<b^*$, we take $b=0.045$ and present the condensate curves in the left panel of Figure~\ref{3}, where we can see that as the temperature $T$ gradually reduces from the region dominated by the normal phase, the s-wave order condensates before the p-wave order. Later the condensate of the p-wave order grows while the condensate of the s-wave order is depressed. After reaching a maximum, the p-wave condensate decrease while the s-wave condensate grows back and finally reaches the value of the single condensate solution. Such a non-monotonic curve of the p-wave condensate in the s+p phase form a shape of letter ``n'' and this case is called the ``n-type''. 
For the second case with $b^*<b<b^o$, we take $b=0.055$ and the condensate curves are presented in the middle panel of Figure~\ref{3}. We can see that in this case, the p-wave condensate will reach the value of the pure p-wave solution while the s-wave condensate vanishes, and the s+p phase is split into two sections of the ``X-type''.
The third case is above the quadruple point with $b>b^o$ and we choose $b=0.1$ to show the condensate in the right penal of Figure~\ref{3}, where the p-wave order emerges first in the high temperature region instead of the s-wave order, and the s-wave condensate dominates the low temperature region. Only one ``X-type'' region of the s+p phase exists between the two single condensate phases.
\begin{figure}
\includegraphics[width=0.3\columnwidth]{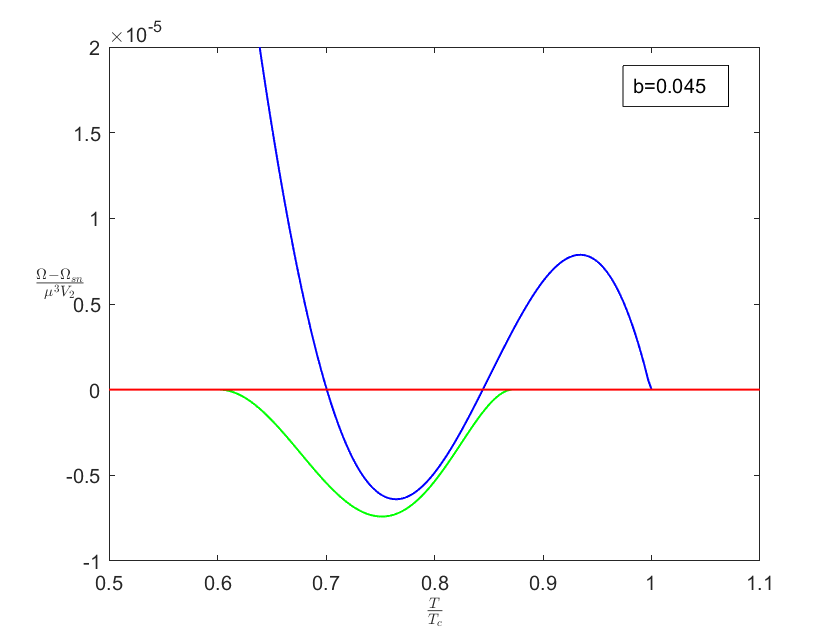}
\includegraphics[width=0.3\columnwidth]{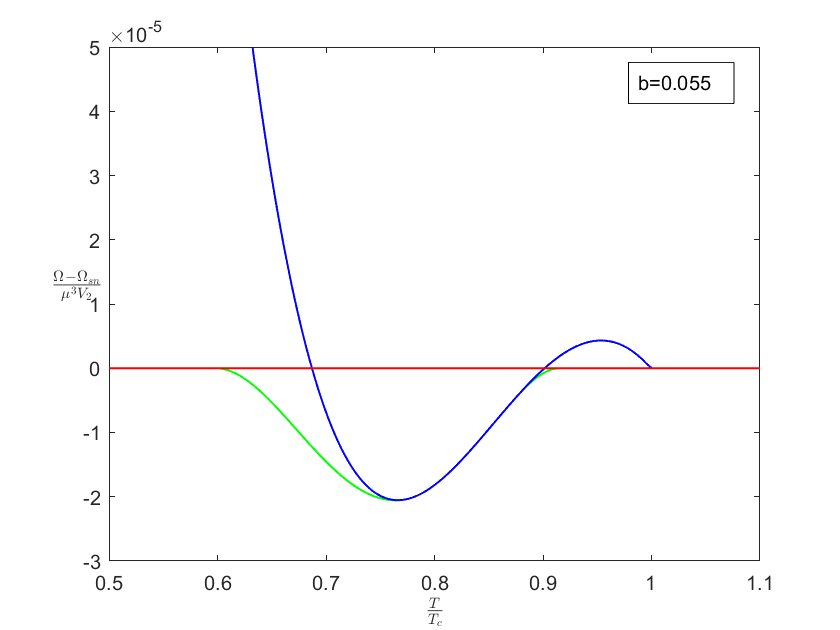}
\includegraphics[width=0.3\columnwidth]{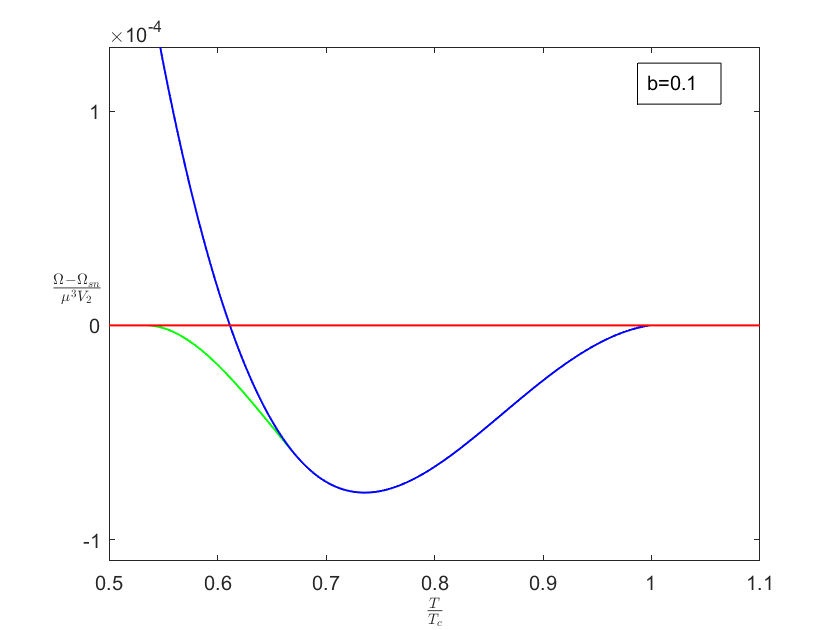}
\caption{The relative value of the grand potential with respect to the s-wave solutions for the three typical cases with $b=0.045$ (Left), $0.055$ (Middle) and $0.1$ (Right). The curves for the p-wave and s+p solutions and colored red and green respectively.
   }\label{4}
\end{figure}

To better understand the above phase transitions, we plot the relative value of the grand potential of the single condensate p-wave solution and s+p solution with respect to the single condensate s-wave solution $\Omega-\Omega_s$ for the three typical cases in Figure~\ref{4}. In the third case where the p-wave order condensates before the s-wave order, we plot the relative value with respect to the normal phase instead in the region above the critical temperature of the single condensate s-wave phase. Therefore, we use a combined notation $\Omega-\Omega_{sn}$ to denote this relative value. In these plots, the red, blue, and green curves represent the grand potential curves for the s-wave solutions, the p-wave solutions, and the s+p solutions, respectively. From the grand potential curves in the three typical cases, it is shown that the grand potential of the s+p solutions is always the lowest in its existence region.

\begin{figure}
\includegraphics[width=0.32\columnwidth]{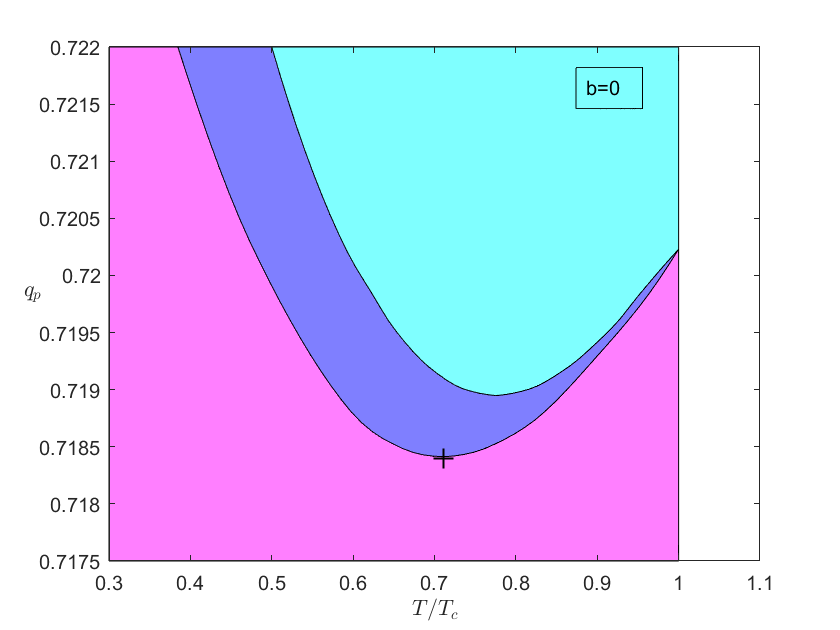}
\includegraphics[width=0.32\columnwidth]{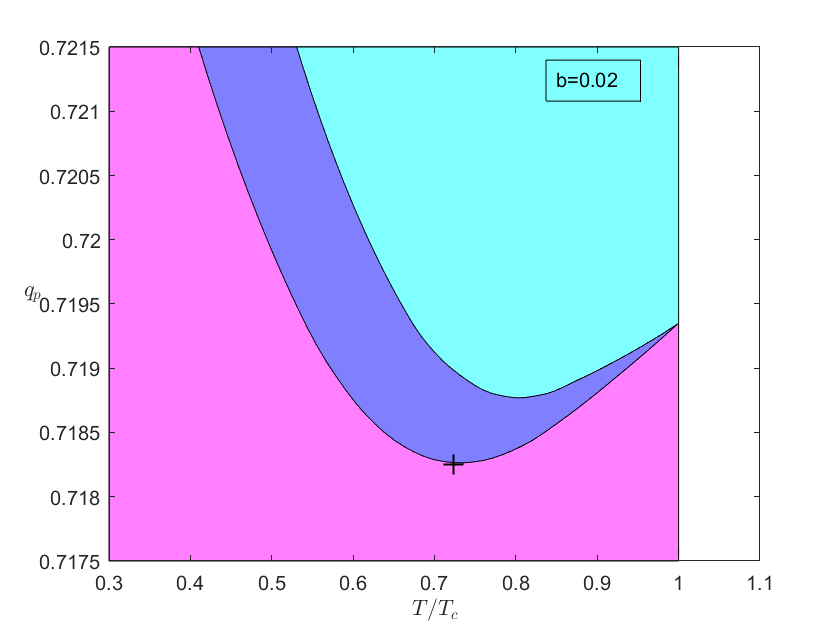}
\includegraphics[width=0.32\columnwidth]{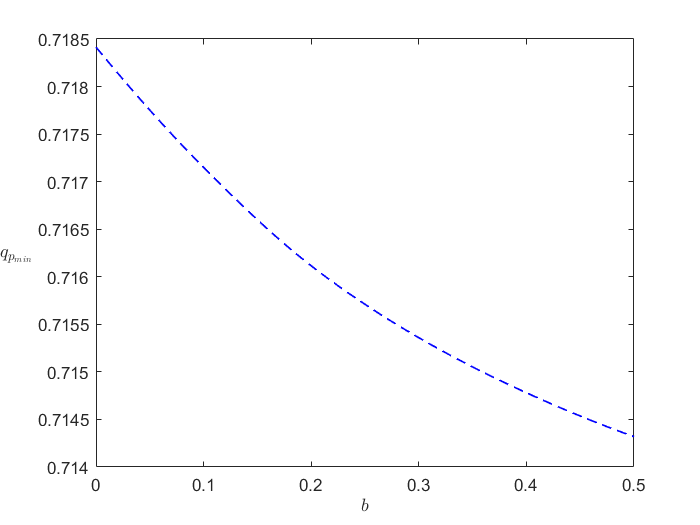}
\caption{The $q_p-T$ phase diagrams for $b=0$ (Left), $b=0.02$ (Middle), and the ${q_p}_{min}-b$ relation (Right). In the left and middle panels, the white, magenta, cyan and blue regions represent the normal phase, the s-wave phase, the p-wave phase and the s+p phase, respectively. The symbol $+$ marked the minimum charge ${q_p}_{min}$ that admits the coexistence s+p solutions, whose dependence on $b$ is plotted in the right panel.}
\label{5}
\end{figure}

To show the more complete phase structure involving the s-wave and p-wave orders, we also consider more general values of the ratio $q_p/q_s=q_p$. We choose two typical values for the nonlinear parameter $b$ and plot the $q_p-T$ phase diagrams in the left and middle panels of Figure~\ref{5}. We see that both the two phase diagrams are divided into four regions dominated by the normal phase (white), the s-wave phase (magenta), the p-wave phase (cyan), and the s+p phase (blue), respectively. The topology of these two phase diagrams are the same as the $b-T$ phase diagram shown in Figure~\ref{2}. From the two $q_p-T$ phase diagrams, it is clear that even at a fixed value of $b$, it seems always possible to get the various phase transitions as shown in Figure~\ref{3} by tuning the value of $q_p$. However, the qualitative minimal value of $q_p$ for a stable section of the s+p coexistent phase is different, and we further plot the dependence of $q_{p_{min}}$ on $b$ in the right panel of Figure~\ref{5}, which suggests that with a larger value of the nonlinear parameter $b$, a smaller $q_p$ is sufficient for the emergence of the multi-condensate s+p solutions.
%%%
%%%
\section{The superconducting charge density}\label{sec4}
The superconducting charge density is usually recognized as the accumulated charge outside the horizon when the charged hair is formatted from the bulk point of view. However, nonlinear electrodynamics makes things different.

The gauge field near the AdS boundary still expands
\begin{align}
\Phi(r)=\mu-\frac{\rho_t}{r}+\cdots,
\end{align}
where $\rho_t$ is the total charge density. Near the black hole horizon $r=r_h$, the regularity condition still requires $\Phi(r_h)=0$, and the near horizon expansion of the gauge field is
\begin{align}
\Phi(r)=\Phi_1(r-r_h)+O((r-r_h)^2)~.
\end{align}
From the bulk point of view, the Gauss's law indicate that the charge density stored inside the horizon is
\begin{align}
\rho_h=\Phi'(r_h)=\Phi_1~,
\end{align}
Which is understand as the normal charge density on the boundary field theory. Consequently, the superconducting charge density is defined as the total charge density minus the normal charge density
\begin{align}
\rho_o=\rho_t-\rho_h~.\label{rho_o}
\end{align}

The superconducting charge density should be zero in the normal phase, which is indeed followed in holographic superconductor models coupled to linear electro magnetic fields with the above definition~\cite{Gubser:2008wv, Nie:2013sda}. However, while considering the nonlinear electrodynamics, the charge accumulates outside the horizon according to the effective self-interaction introduced by the nonlinear term from the bulk point of view. Therefore, the value of $\rho_o$ from the above formula (\ref{rho_o}) is no longer zero in the normal phase and we should consider new appropriate definition for the superconducting charge density.

One simple redefinition of the superconducting charge density could be taken as the relative value of $\rho_o$ with respect to the normal solution
\begin{align}
\rho_{\text{s}}=\rho_{\text{o}}-\rho_{\text{o}}^{n}~,
\end{align}
where $\rho_{\text{o}}^{n}$ is the value of $\rho_o$ for the normal solution at the same temperature.

\begin{figure}
\includegraphics[width=0.3\columnwidth]{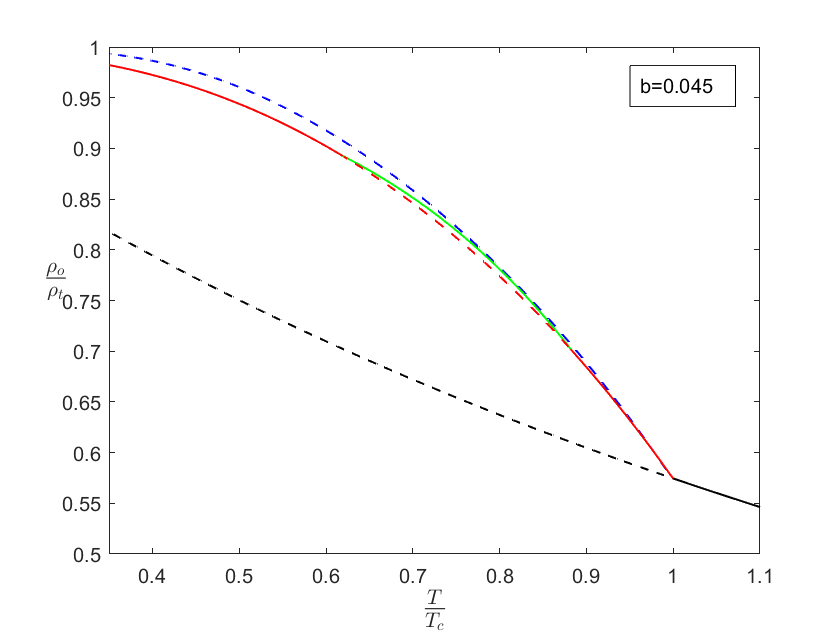} \includegraphics[width=0.3\columnwidth]{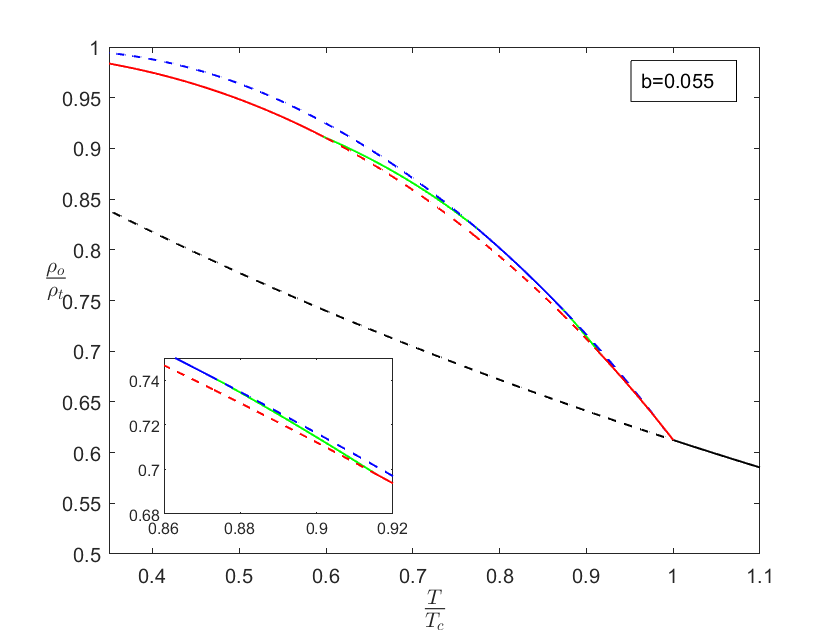}   \includegraphics[width=0.3\columnwidth]{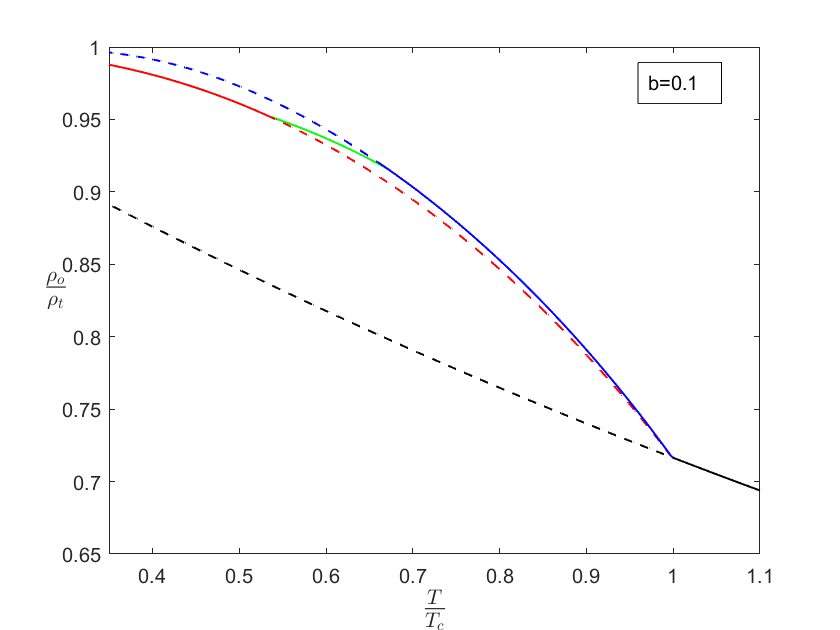}
\caption{The ratio of $\rho_o/\rho_t$ for the various solutions with $q_p=0.7183$ and the three typical values for the nonlinear parameter $b=0.045$ (Left), $0.055$ (Middle) and $0.1$ (Right). The black, red, blue, and green lines represent the normal solutions, the s-wave solutions, the p-wave solutions, and the coexisting s+p solution, respectively. Solid lines denote the most stable solutions, while dashed lines denote the unstable solutions.
}\label{6}
\end{figure}

\begin{figure}
\includegraphics[width=0.3\columnwidth]{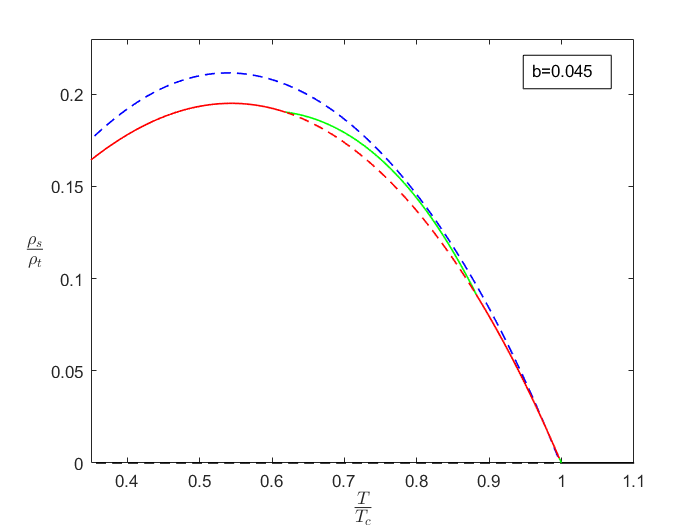}
\includegraphics[width=0.3\columnwidth]{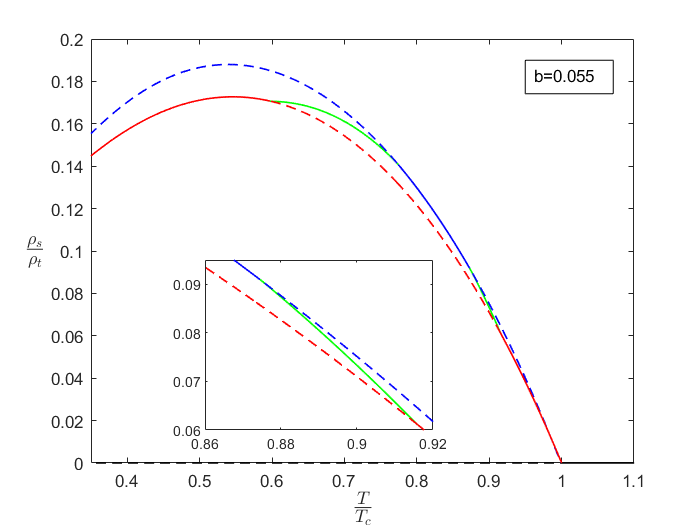}
\includegraphics[width=0.3\columnwidth]{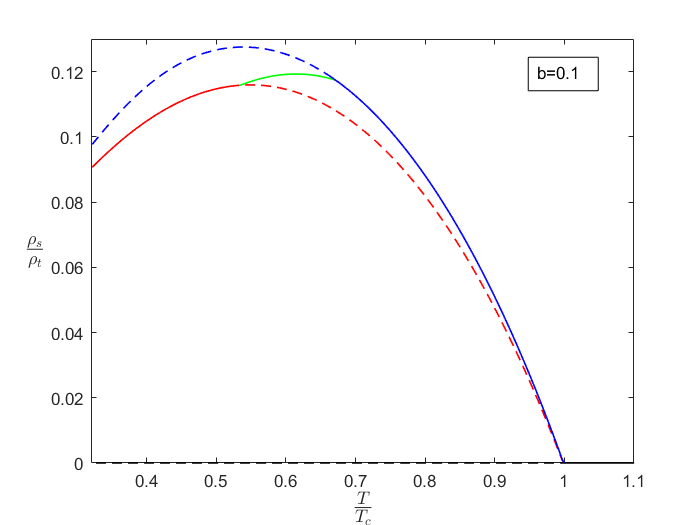}
\caption{The ratio of $\rho_s/\rho_t$ for the various solutions with $q_p=0.7183$ and the three typical values for the nonlinear parameter $b=0.045$ (Left), $0.055$ (Middle) and $0.1$ (Right). The black, red, blue, and green lines represent the normal solutions, the s-wave solutions, the p-wave solutions, and the coexisting s+p solution, respectively. Solid lines denote the most stable solutions, while dashed lines denote the unstable solutions.}\label{7}
\end{figure}

In Figure~\ref{6}, we plot the ratio of $\rho_o/\rho_t$ for the various solutions with $q_p=0.7183$ and the three typical values for the nonlinear parameter $b=0.045$ (Left), $0.055$ (Middle) and $0.1$ (Right), respectively. The black, red, blue and green lines represent the normal solutions, the s-wave solutions, the p-wave solutions, and the s+p solutions, respectively. Solid lines are used to denote the most stable solutions while the dashed lines indicate the unstable sections. It is clear that with the usual definition of the superconducting charge density, $\rho_o$ is not zero in the normal solutions. We further plot the ratio with our newly defined superconducting charge density $\rho_{\text{s}}/\rho_t$ for the various solutions for the same three typical cases in Figure~\ref{7}. We can see that the superconducting charge density for the normal solutions are now zero. However, the superconducting charge density curves exhibit a non-monotonic dependence on temperature. As the temperature decreases, the ratio $\rho_s/\rho_t$ increases below the critical point, but decreases in the low temperature region, which is quite different with the monotonic behavior for the holographic superconductors coupled to linear electrodynamics.
\section{The optical conductivity}\label{sec5}
Since the nonlinear electrodynamics brings additional charge outside the horizon, it is nature to further examine the optical conductivity. The linear perturbations of the gauge field component $\delta A_y$ decoupled from other modes and can be assumed to be $\delta A_y=A_ye^{-i\omega t}$ due to the time translation symmetry. The  linearized equation for the perturbation $A_y$ is
\begin{align}
	A_y''(r)+\left[\frac{f'(r)}{f(r)}+\frac{2b\Phi'(r)\Phi''(r)}{1+b\Phi'^2(r)}\right] A_y'(r)+\left[\frac{\omega^2}{f^2(r)}-\frac{2\Psi_p^2(r)+2r^2\Psi_s^2(r)}{r^2f(r) (1+b\Phi'^2(r))}\right]A_y(r)=0~.
\end{align}
Ingoing boundary condition is applied at the event horizon
\begin{align}
	 A_y(r)=(r-r_h)^{-i\omega L^2/3r_h}\left(1+A_{y1}(r-r_h)+A_{y2}(r-r_h)^2+\cdots\right) ~.
\end{align}
while the asymptotic expansion of $A_y$ near the boundary is 
\begin{align}
	A_y(r)=A^{(0)}+\frac{A^{(1)}}{r}+\cdots~.
\end{align}
The AdS/CFT correspondence tells us that the dual source and the current expectation value are represented by $A^{(0)}$ and $A^{(1)}$, respectively. According to Ohm’s law, we obtain the conductivity as
\begin{align}
\sigma(\omega)=-\frac{iA^{(1)}}{\omega A^{(0)}}~.\end{align}

\begin{figure}
\includegraphics[width=0.33\columnwidth]{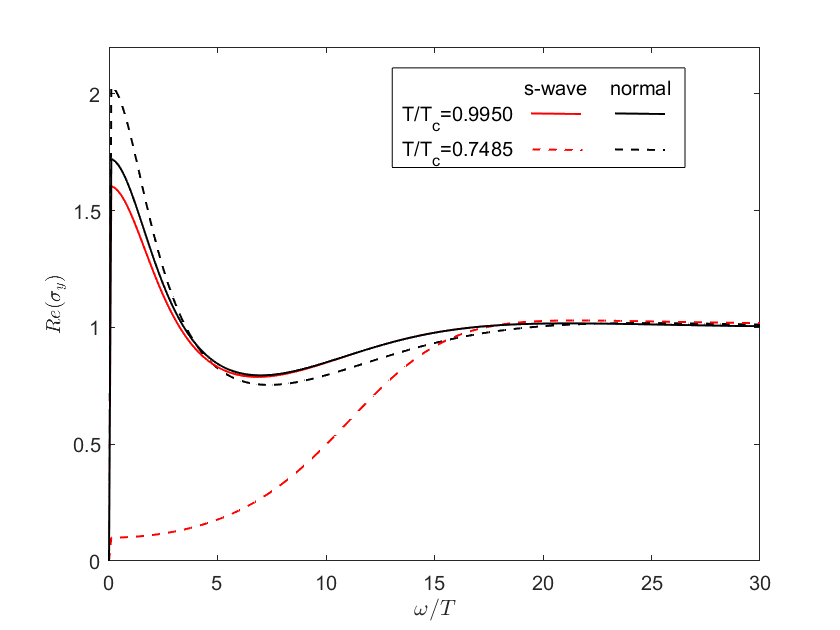}
\includegraphics[width=0.33\columnwidth]{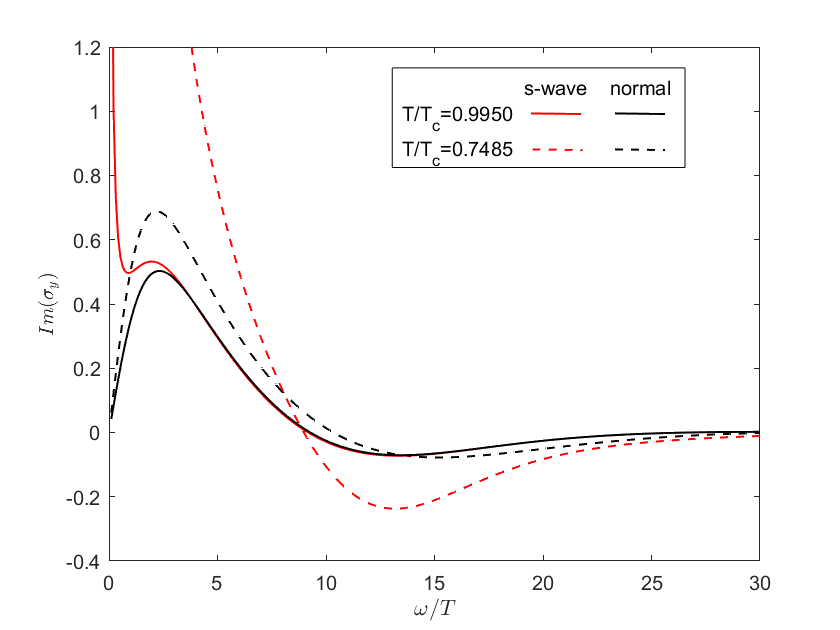}
\includegraphics[width=0.33\columnwidth]{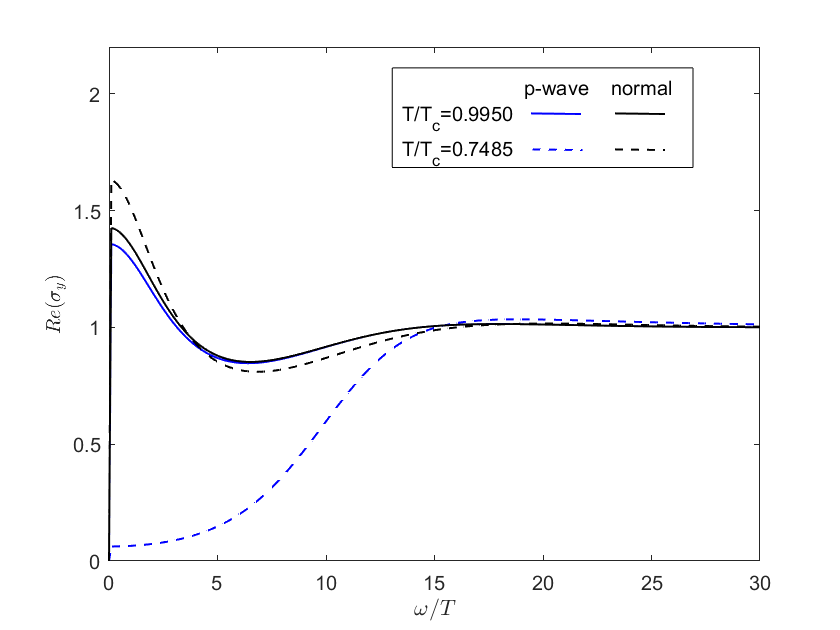}
\includegraphics[width=0.33\columnwidth]{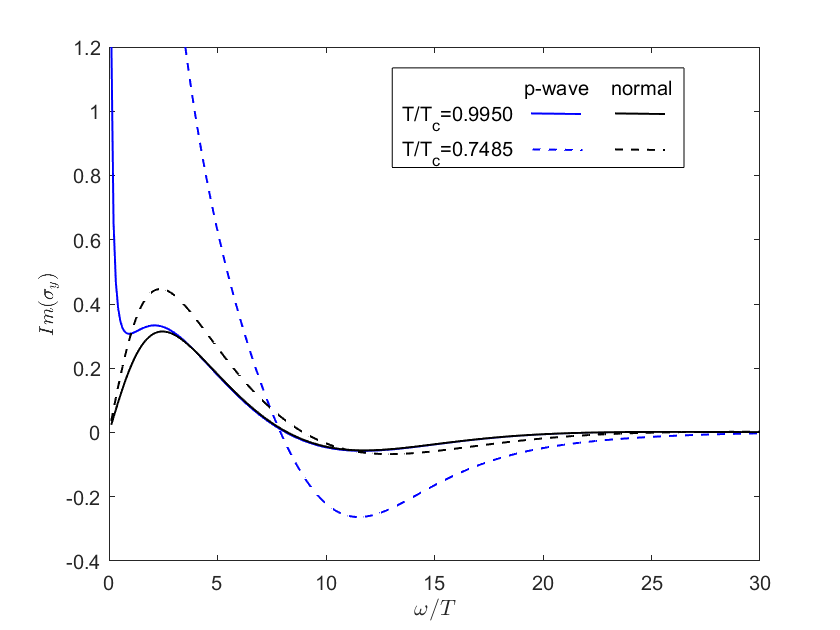}
\caption{The real (Left) and imaginary (Right) parts of the optical conductivity for the normal solutions as well as the single condensate s-wave (Top) and p-wave (Bottom) solutions with $b=0$ and $q_p=q_s=1$.
The black, red and blue lines represent the normal solutions, the s-wave solutions and the p-wave solutions, respectively. The solid lines indicate the cases with $T/T_c=0.9950$ and the dashed lines indicate the cases with $T/T_c=0.7485$.
}\label{8}
\end{figure}

With the above standard schedule, we are able to investigate the optical conductivity for the normal solutions as well as the superconducting solutions with various condensates.
We plot the real and imaginary parts of the optical conductivity for the normal solutions as well as the single condensate s-wave and p-wave solutions with the nonlinear parameter $b=0$ and $q_p=q_s=1$ in Figure~\ref{8}. We use the black lines to indicate the results of the normal solutions, and we can see that the optical conductivity in the normal solutions are no longer constants as in the case of linear electrodynamics. Nontrivial configurations are presented in the low frequency region, which is expected to be related to the additional charge accumulated outside the horizon from the bulk point of view.

Since the condensed solutions near the critical points are still close to the normal solutions, the nontrivial profile is possible to be inherited in the superconducting solutions close to the critical points. The red and blue solid lines in Figure~\ref{8} show the results for the single condensate s-wave and p-wave solutions at $T/T_c=0.9950$, which is very close to the critical points. We can see that both the real and imaginary parts of the optical conductivity in these superconducting solutions are very similar to the curves for normal solutions at the same temperature.

\begin{figure}
\includegraphics[width=0.3\columnwidth]{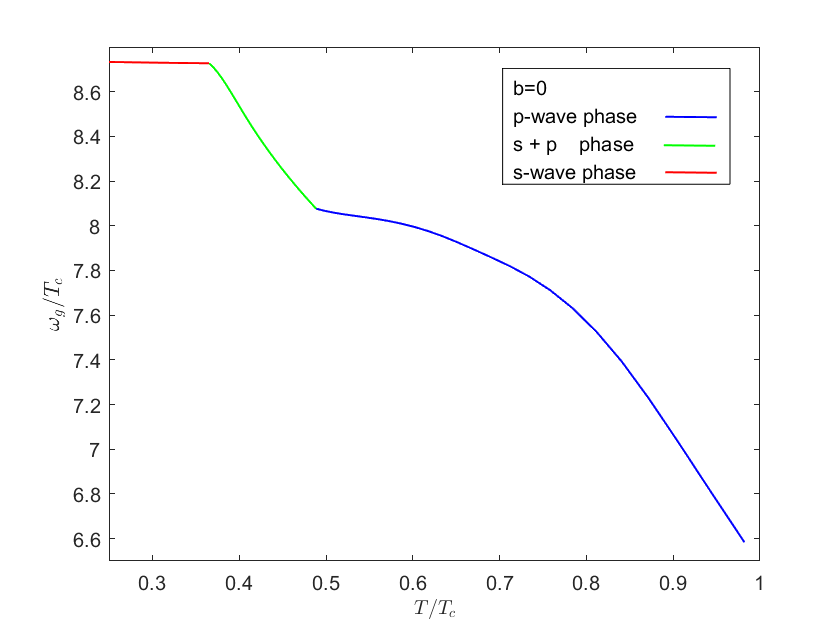}
\includegraphics[width=0.3\columnwidth]{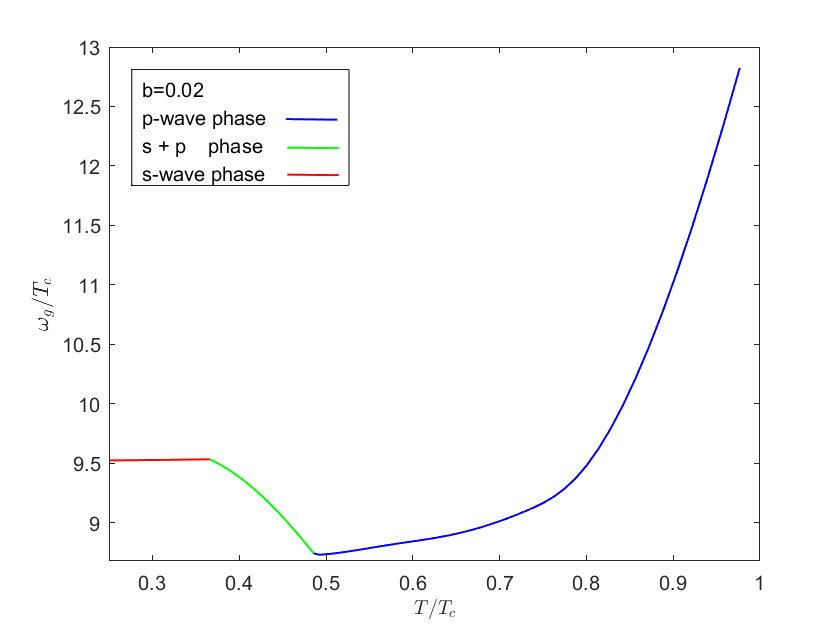}
\caption{The dependence of the gap frequency $\omega_g/T_c$ on the temperature $T/T_c$ for $b=0$ (Left) and $b=0.002$ (Right) with $q_p=0.7225$. The red, blue, and green lines represent the s-wave solutions, the p-wave solutions, and the s+p solutions, respectively.
}\label{9}
\end{figure}

The frequency gap in the superconducting phase is usually read from the location of the minimum in the imaginary part of the optical conductivity. Therefore, the new feature in the optical conductivity close to the critical point will confuse the location of the gap frequency. As an example, we plot the gap frequency $\omega_g/T_c$ located by the minimum of the imaginary part of the optical conductivity in Figure~\ref{9}, where $q_p$ is set to $0.7225$. The left panel is for the case of linear electrodynamics with $b=0$ and the right panel is for the case with $b=0.02$. Comparing the two panels, we can see that while the gap frequency in the two cases are similar in the low temperature region, the results near the critical point is quite different. In the linear case with $b=0$, the gap frequency show negative dependence on the temperature near the critical point, while the nonlinear case with $b=0.02$ show negative dependence. This drastic change is explained by the inherited behavior of optical conducntivity in the superconducting solutions near the critical points.
\section{Conclusions and discussions} \label{sec6}
In this paper, we explored a holographic s+p superconductor model coupled to nonlinear electrodynamics in the probe limit. The study focused on the effects of the nonlinear parameter $b$ on the phase transitions involving the two orders, the superconducting charge density as well as the optical conductivity. For the single condensate s-wave and p-wave solutions, the critical temperature $T_c/\mu$ increases along the increasing of the nonlinear parameter $b$. We calculated the condensates of various phases and plot the grand potential curves to confirm the stability. It is found that the grand potential curves for both the single condensate s-wave and p-wave solutions rise up as the nonlinear parameter $b$ increases. The results consistently  indicate that a larger $b$ reduces the stability of the single condensate s-wave and p-wave solutions. We also further plot the $b-T$ phase diagram as well as $q_p-T$ phase diagrams to present the more complete phase structures including the multi-condensate s+p phases.

The new features come in the study of superconducting charge density, where the usual definition of superconducting charge density $\rho_o$ get a non zero value even in the normal solutions. In order to solve this problem, we give a new definition for the superconducting charge density $\rho_s$ as the relative value of $\rho_o$ in the supercondcuting solutions with respect to $\rho_o$ in the normal solutions at the same critical point. With this new definition, the non-zero problem for the normal solutions is solved and the superconducting charge density show a non monotonic dependence on temperature, which is different to the results in holographic models coupled to linear electrodynamics.

Furthermore, we studied the optical conductivity of normal solutions and the superconducting solutions in more detail. Due to the nonlinear electrodynamics, the optical conductivity for the normal phase shows nontrivial profile in the low frequency region, and develops a minimum in the imaginary part. This profile is inherited in the superconducting solutions close to the critical points, and confuses the location of the gap frequency. As a result, if we still locate the gap frequency as the imaginary part of the optical conductivity, the dependence of the gap frequency on temperature changes drastically when the nonlinear parameter $b$ is turned on.

The above strange features in the superconducting charge density as well as the optical conductivity result from the additional charge accumulated outside the horizon due to the nonlinear self interaction introduced by the nonlinear electrodynamics from the bulk point of view. Further attentions should be paid to these new features and the further investigations for new appropriate definitions and calculations in the holographic models coupled to nonlinear electrodynamics. 

In this work, we only considered the model in the probe limit. Future studies should further examine the back-reaction of the matter fields on the metric. Moreover, it is interesting to introduce the perturbations of $A_x$ and the coupled components to investigate the anisotropic behavior of conductivity in the p-wave and s+p phases. We hope to see studies on these interesting issues in future.
\section*{Acknowledgements}
This work was supported by the National Natural Science Foundation of China (Grant Nos. 12575054 and 11965013). ZYN is partially supported by Yunnan High-level Talent Training Support Plan Young \& Elite Talents Project (Grant No. YNWR-QNBJ-2018-181).
\bibliographystyle{apsrev4-1}
\bibliography{reference}
\end{document}